\documentclass[utf8]{FrontiersinHarvard} 

\usepackage{url}
\usepackage{hyperref}
\usepackage{lineno}
\usepackage{microtype}
\usepackage{amssymb}
\usepackage{algorithmic}
\usepackage{amsmath}
\usepackage[onehalfspacing]{setspace}

\def\keyFont{\fontsize{8}{11}\helveticabold }
\def\firstAuthorLast{Akhaury {et~al.}} 
\def\Authors{Utsav Akhaury\,$^{1,*}$, Jean-Luc Starck\,$^{2}$, Pascale Jablonka\,$^{1}$, Frédéric Courbin\,$^{1}$ and Kevin Michalewicz\,$^{1}$}
\def\Address{$^{1}$ Laboratory of Astrophysics, École Polytechnique Fédérale de Lausanne (EPFL), CH-1015 Lausanne, Switzerland. \\
$^{2}$AIM, CEA, CNRS, Université Paris-Saclay, Université de Paris, F-91191, Gif-sur-Yvette, France. }
\def\corrAuthor{Utsav Akhaury}

\def\corrEmail{utsav.akhaury@epfl.ch}

\begin{document}
\onecolumn
\firstpage{1}

\title[DL-based galaxy image deconvolution]{Deep Learning-based galaxy image deconvolution} 

\author[\firstAuthorLast ]{\Authors} 
\address{} 
\correspondance{} 

\extraAuth{}

\maketitle

\begin{abstract}

With the onset of large-scale astronomical surveys capturing millions of images, there is an increasing need to develop fast and accurate deconvolution algorithms that generalize well to different images. A powerful and accessible deconvolution method would allow for the reconstruction of a cleaner estimation of the sky. The deconvolved images would be helpful to perform photometric measurements to help make progress in the fields of galaxy formation and evolution. We propose a new deconvolution method based on the Learnlet transform. Eventually, we investigate and compare the performance of different Unet architectures and Learnlet for image deconvolution in the astrophysical domain by following a two-step approach: a Tikhonov deconvolution with a closed-form solution, followed by post-processing with a neural network. To generate our training dataset, we extract HST cutouts from the CANDELS survey in the F606W filter (V-band) and corrupt these images to simulate their blurred-noisy versions. Our numerical results based on these simulations show a detailed comparison between the considered methods for different noise levels.

\tiny
 \keyFont{ \section{Keywords:} Deconvolution, Denoising, Image Processing, Deep Learning, Inverse problem, Regularization} 
\end{abstract}

\section{Introduction}

In the upcoming decade, large telescopes such as the Vera C. Rubin Observatory (LSST) and Euclid will offer a broader view of the universe by capturing several images. Both these telescopes would be covering a huge part of the sky, thus giving us access to a wide variety of objects, and capturing images in various frequency bands. Each frequency band captures certain unique information, which is beneficial for tracing the constituent components of galaxies. In addition, these surveys aim to go deeper - thus capturing fainter objects, and to higher redshifts - thus capturing very distant objects. A huge variety of galaxy images would help us better understand their origin and evolution. From an astrophysical point of view, the ultimate aim is to translate the information carried by these images and derive physical inferences (for eg: obtaining the metallicity, morphology, or flux of different galaxies). Unfortunately, imperfections are generated by every image acquisition system. Usually, a blurring effect is introduced in the images, which is modeled by a Point Spread Function (PSF) and is considered to be space-invariant. Moreover, the sensor variations introduce noise in the images, which is usually additive, white, and Gaussian. As such, there is a dire need to develop fast and accurate deconvolution algorithms that generalize well to different images. In addition, bad pixels or cosmic ray hits also need to be taken into account. Generally, they are relatively easily identifiable, and a good solution would be to apply an in-painting method to replace the bad pixel values by reasonable ones. In this paper, we assume that this pre-processing step has been carried out, and that we can directly deal with the clean images. A powerful and accessible deconvolution method would allow for the reconstruction of a cleaner estimation of the sky. Deconvolution is useful for a broad range of applications, such as galaxy morphology studies, substructure identification in galaxies, bulge/disk separation, etc. Deconvolution is also very important for comparing two images at different resolutions. Hence, technical development in the field of image processing is essential to answering the fundamental questions in astrophysics.

By using the least-squares method, one can partially reconstruct the image; however, the solution oscillates while solving the equation since the problem is ill-conditioned. Broadly speaking, it belongs to the family of ill-posed problems that could alternatively be handled using regularization. Due to these effects, simply minimizing the Mean Squared Error (MSE) between the observation and the reconstruction does not lead us to proper convergence. To tackle this, constraints associated with the signal's energy, such as its derivatives \citep{bertero1998introduction}, total variation \citep{chambolle2010introduction,rudin1992nonlinear} or sparsity \citep{farrens2017,starck2015sparse}, could be added. These routinely used methods are apt for solutions optimizing the MSE. For such inverse problems, sparse wavelet regularization using the $\ell_0$ or $\ell_1$ norm has remained the routinely used approach for astrophysical image deconvolution. It has led to striking results, such as an improvement in resolution by a factor of four in the Cygnus-A radio image reconstruction compared to the standard CLEAN algorithm \citep{garsden2015}. Sparsity could be considered a weak prior on the distribution of the wavelet coefficients of the solution. This comes from the fact that most images could be represented in a more compressed manner in the wavelet domain. The advancement in deep learning over the recent past has presented encouraging outcomes in the field of deconvolution \citep{xu2014dcnn}. Within the astrophysics community, deep learning approaches have been introduced to carry out model fitting that could be considered as a parametric deconvolution \citep{tuccillo2018deepfitting}. Particularly, Unets \citep{ronnenberger2015Unet} have gathered tremendous attention due to their effectual performance, which can be attributed to their highly non-linear processing and the availability of large training datasets. Based on Unet, \citet{sureau2020} developed the Tikhonet method for deconvolving galaxy images in the optical domain. Tikhonet is a two-step deep learning-approach, the first being a Tikhonov deconvolution, i.e., with a standard quadratic regularization, followed by a neural network denoising using a 4-scale \emph{XDense Unet}. Giving the Tikhonov deconvolution as an input to the network , the training aims at minimizing the MSE between the reconstruction and the ground truth image. It was shown that Tikhonet surpassed sparse regularization-based methods in terms of the MSE and a shape criterion, where a measure of the galaxy ellipticity was used to encode its shape \citep{sureau2020}. 

While the multiple instances of non-linearity are inherent to artificial neural networks, the idea of learning with training data could even be extended to methods that exploit sparsity. 
Recently, \citet{ramzi:hal-03346892} presented a novel architecture, termed Learnlet, that preserves the properties of sparsity-based approaches (for example linear decomposition and reconstruction steps, good generalization properties, exact reconstruction) while simultaneously capturing the prowess of neural networks. The Learnlet decomposition aims at learning a filter bank in a denoising setting with backpropagation and gradient descent. Learnlets were originally proposed for denoising and have been shown to have better generalization properties than Unets \citep{ramzi:hal-03346892}, while Unets showed better performances. In this paper, we investigate if the Learnlet neural network could also be a good alternative to Unet for deconvolution. We propose a new deconvolution approach based on the Learnlet decomposition, using the same two-step approach as in \citet{sureau2020}, but by substituting the Unet denoiser by Learnlet. We compare our results on images extracted from the CANDELS survey \citep{CANDELS, CANDELS_HST}. In section \ref{dldeconv}, we introduce the deconvolution problem and the deep learning-methods that have been developed to tackle them. In section \ref{learnlet}, we discuss the concept of Learnlet decomposition as introduced by \citet{ramzi:hal-03346892}, and extend the idea to use the network for image deconvolution. We detail out the process of generating our dataset and perfoming the numerical experiments in section \ref{exp}, and in section \ref{results}, we demonstrate the results obtained for these experiments. Finally, in section \ref{conclusion}, we conclude our work.

\section{Deep Learning-based Deconvolution}
\label{dldeconv}

\subsection{The Deconvolution Problem}

Let $\mathbf{y}\in\mathbb{R}^{n\times n}$ be the observed image and $\mathbf{h}\in\mathbb{R}^{n\times n}$ be the PSF. The observed image can be modelled as
\begin{equation}
 \label{eq:invprob}
  \mathbf{y} = \mathbf{h} \ast \mathbf{x_t} + \mathbf{\eta}
\end{equation}
where $\mathbf{x_t} \in \mathbb{R}^{n \times n}$ denotes the target image, $\mathbf{*}$ denotes the convolution operation, and $\mathbf{\eta} \in\mathbb{R}^{n\times n}$ denotes additive Gaussian noise. Our aim is to recover the ground truth image $\mathbf{x_t}$. One could partly recover an estimate $\hat{\mathbf{x}}$ of $\mathbf{x_t}$ by the help of the least-squares method. However, the solution would oscillate because equation \ref{eq:invprob} is ill-conditioned. More broadly, it falls under the category of an ill-posed problem, which could be handled by regularization \citep{bertero1998introduction}. 

Convolutional Neural Network architectures like Unet \citep{ronnenberger2015Unet} have shown to be very efficient in image noise removal, as seen in \cite{gurrola2021Unet}. The Unet was originally developed for biomedical image segmentation. Since then, it has been found to be relevant to many other imaging problems, not just segmentation. Meanwhile, denoising methods based on wavelets are no longer the state-of-the-art. However, they have theoretical guarantees (as seen in \cite{donoho1995noising}). Unets consist of a multi-scale approach similar to wavelets, which allows the signal to be analyzed at multiple resolutions. The prime difference comes from the usage of non-linearities at various steps. Wavelets include only a single non-linearity step (wavelet shrinkage) when used for denoising. Contrastingly, Unets rely on several ReLU and max-pooling steps. The denoising analysis in Unets becomes very complicated due to such chained non-linearities. Particularly, it is tough to understand how a network that is trained on one kind of noise could be used for other kinds of noise. Moreover, a few works \citep{gottschling2020troublesome} even demonstrate that deep learning-based methods are unable to recover features that their classical counterparts can. 

For a ground truth image $\mathbf{x} \in \mathbb{R}^{n \times n}$, let $\mathbf{\tilde{x}} = \mathbf{x} + \mathbf{\epsilon}$ be its corrupted version generated by adding Gaussian noise $\mathbf{\epsilon} \sim \mathcal{N}(\mathbf{0}, \mathbf{C_{n \times n}})$ with a known covariance $ \mathbf{C_{n \times n}}$. For a denoiser $\mathbf{N}_{\theta}$, the general optimization problem can be represented as:

\begin{equation}
    {\text{argmin }}  \mathbb{E}_{\mathbf{x}} \left[ \|\mathbf{x} - \mathbf{N}_{\theta}(\mathbf{\tilde{x}}) \|_2^2 \right]
    \label{eq:learning}
\end{equation}

\subsection{The Tikhonet Solution}
\label{tikhonet}

Tikhonet is a two-step deep learning-based deconvolution method. The first step involves deconvolving the input image with a Tikhonov filter based on a quadratic regularization. Let $\mathbf{H}\in\mathbb{R}^{n^2 \times n^2}$ be the circulant matrix associated with the convolution operator $\mathbf{h}$. Then, the Tikhonov solution of equation \ref{eq:invprob} can be written as:
\begin{equation}
    \label{eq:tikhonov}
    \mathbf{\hat{x} = \left(H^\top H+\lambda \Gamma^\top \Gamma\right)^{-1}H^\top y\quad},
\end{equation}
where $\mathbf{\Gamma}\in\mathbb{R}^{n^2 \times n^2}$ corresponds to the linear Tikhonov filter which is set to a Laplacian high-pass filter to penalize high frequencies, and $\mathbf{\lambda}\in\mathbb{R}_+$ corresponds to the regularization weight, whose appropriate value for every image is found by Stein’s Unbiased Risk Estimate (SURE) minimization \citep{sureau2020}. The Tikhonov step results in a deconvolved image that contains correlated additive noise which can be removed in a following step by a four-scale \emph{XDense Unet}. The training is aimed to make the network learn the mapping from the Tikhonov output $\hat{\mathbf{x}}$ to the ground truth image $\mathbf{x_t}$ while minimizing the MSE (equation \ref{eq:learning}). 

\begin{figure}[h!]
\begin{center}
\includegraphics[width=\columnwidth]{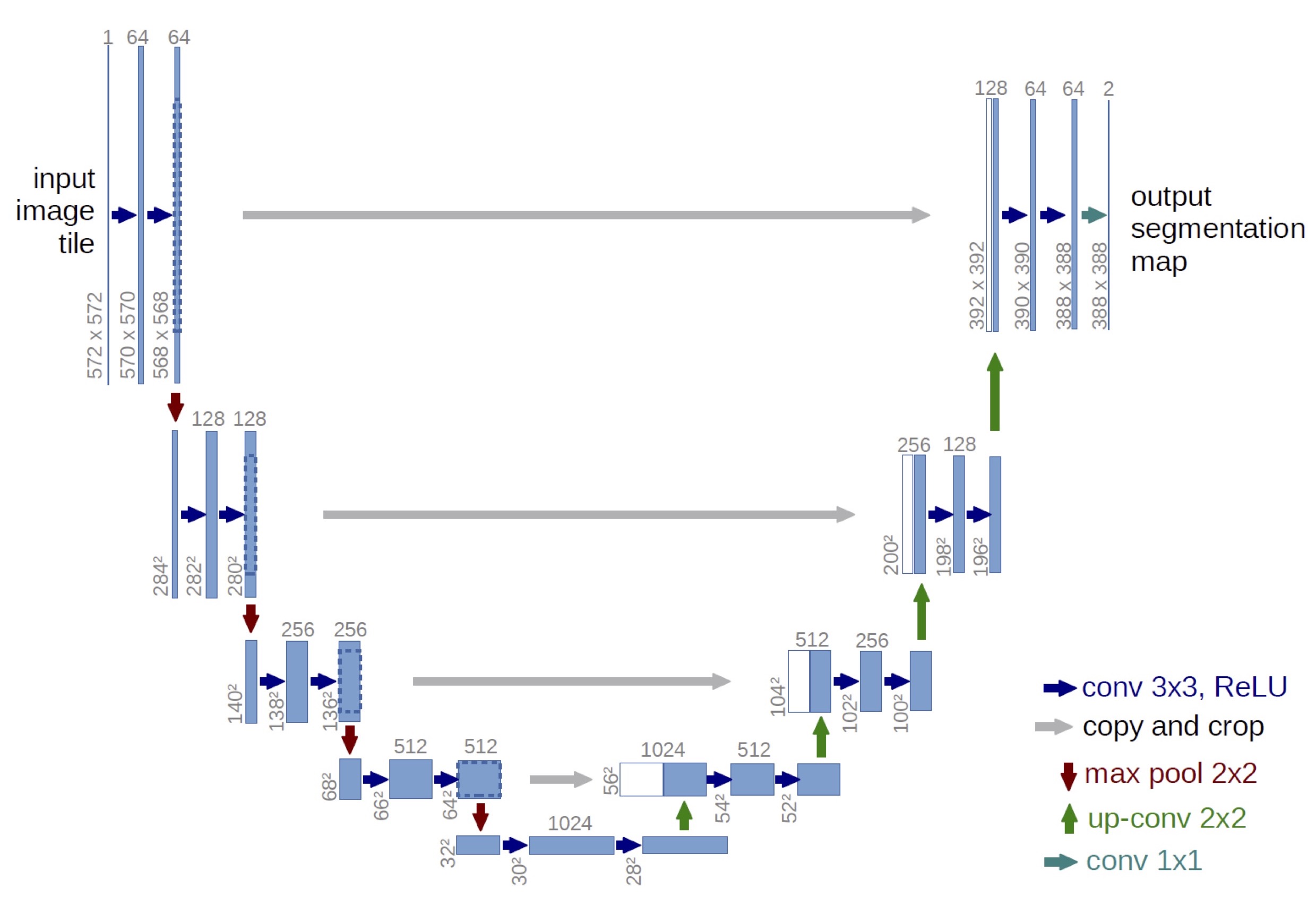}
\end{center}
\caption{\label{fig:Unet} The global Unet architecture. Note that the input dimensions are $128 \times 128$ in our case. The number of channels for each feature map is indicated on the top of the blue rectangles. Here, there are 64 base filters. The different scales in this model help analyze the images at multiple resolutions. Credit: Ronneberger et al. (2015).}
\end{figure}

The model architecture heavily impacts the denoising performance. It should have some multi-scale processing in order to capture distant correlations, pointing to the usage of a Unet like layout \citep{ronnenberger2015Unet} shown in figure \ref{fig:Unet}, which has already found success in solving inverse problems \citep{jin2017cnn}. Ideally, one should also aim to bring down the trainable parameter count. Considering these factors, the \emph{XDense Unet} inherits the global layout from \citep{jin2017cnn}, but with these alterations \citep{sureau2020}:

\begin{itemize}

    \item At each scale, the convolutional layers were replaced by \textbf{dense blocks} introduced by \cite{huang2017cvpr}, which decrease the parameter count by concatenating feature maps from the previous layers to the current layer's input, and were claimed to help preserve information, enable reusing features, and limit vanishing gradients.
    
    \item 2D convolutions were substituted by \textbf{2D separable convolutions} \citep{chollet2016}, which helped decrease the parameter count in the model by assuming that correlations across feature maps and spatial correlations can be independently captured.
    
    \item The max-pooling step was changed to \textbf{average pooling}, since it led to over-segmentation of the final estimates.
    
    \item The \textbf{skip connection} from the input to the output layer \citep{jin2017cnn} was \textbf{removed} , since it degraded the network performance particularly at low Signal-to-Noise ratio (SNR).
\end{itemize}

The first two alterations notably bring down the parameter count per scale of the Unet, thus increasing the potential number of scales for a given number of trainable parameters. Additionally, we remove the biases from the convolutional layers, as they have been shown to lead to a low generalisation capacity \citep{Mohan2020Robust}. In all, the \emph{XDense Unet} model has $184,301$ trainable parameters. 

\subsection{Advantages and Drawbacks of Tikhonet for Astronomical Image Deconvolution}
\label{tikho_drawbacks}

Tikhonet presents many advantages:
\begin{itemize}
    \item It provides better results than standard techniques usually used in astrophysics \citep{sureau2020}.
    \item It is an extremely fast method working for both optical and radio image deconvolution, therefore well-adapted to the forthcoming big data challenges.
    \item It allows to easily take into account the spatial variation of the PSF. Indeed, we know that the PSF in large surveys such as the Canada-France Imaging Survey (CFIS) or Euclid varies spatially. It was shown by \cite{sureau2020} that Tikhonet  can handle such data by deconvolving each galaxy in the field independently and using the PSF corresponding to the center of the galaxy. The PSF field is generally not known, but can be reconstructed from observed stars present in the image \citep{tobias2021, tobias2022}.
    \item Many galaxies share very similar morphologies. Having a learning process allows to capture these morphologies and improve the deconvolution.
\end{itemize}

However, it also has a few drawbacks:
\begin{itemize}
    \item Generalization properties are not so good. It was shown that Unet does not perform well on image morphologies or noise levels that were not included in the training dataset \citep{ramzi:hal-03346892}.
    \item It is not clear how Unets deal with images containing noise with non-white statistical properties (for eg: Poisson, non-stationary Gaussian noise), while sparsity-based techniques can easily consider different kinds of noise \citep{starck2015sparse}.
    \item There is no theory to support the method.
\end{itemize}

In a denoising framework, the Learnlet decomposition has shown interesting properties that could also be considered for deconvolution.

\subsection{Using a Deeper Unet}
In addition to the compact \emph{XDense Unet}, we also tested the performance of \emph{Unet-64}, a deeper Unet as used in \cite{ramzi:hal-03346892} that has $31,023,940$ trainable parameters (around $170$ times more parameters than \emph{XDense Unet}). The architecture is the same as shown in figure \ref{fig:Unet}. As in the case of Tikhonet, we used the bias-free Unet architecture, as these biases lead to a low generalisation capacity \citep{Mohan2020Robust}. The full Python implementation of the code is publicly available. \footnote{https://github.com/zaccharieramzi/understanding-Unets}.

\section{Learnlet Deconvolution}
\label{learnlet}

\subsection{Learnlets}

Denoising methods based on wavelets are no longer the state-of-the-art, but have theoretical guarantees and are the baseline for other approaches. In cases where guarantees are desired, like medical applications, they are the suitable candidates. Recently, \cite{ramzi:hal-03346892} presented a new network architecture, called Learnlet, which exploits one of the most desirable usefulnesses of deep learning: using gradient descent to improve the expressive power of wavelets, while preserving some interesting wavelet properties like exact reconstruction.

In the presence of White Gaussian Noise (WGN), if $\Sigma$ denotes the set of possible values for the noise standard deviation $\sigma$, $m$ the number of scales, and $\theta = (\mathbf{\theta_s}, \mathbf{\theta_t}, \mathbf{\theta_a}) \in \Theta_m$ a given set of parameters, then Learnlets are defined as the following function $\mathbf{f}_{\theta}$ from $(\mathbb{R}^{n \times n} \times \Sigma)$ to $\mathbb{R}^{n \times n}$:

\begin{equation}
   \mathbf{f}_{\theta}(\mathbf{\tilde{x}}, \sigma) = \mathbf{S}_{\mathbf{\theta_s}} \left( \mathbf{T}_{\mathbf{\theta_t}} \left( \mathbf{A}_{\mathbf{\theta_a}} \left( \mathbf{\tilde{x}} \right) , \sigma \right) \right)
\end{equation}

where

\begin{enumerate}
    \item $\mathbf{A}_{\mathbf{\theta_a}}$ is the analysis function, which is equivalent to the wavelet transform with learned filters:
    \begin{equation}
        \mathbf{A}_{\mathbf{\theta_a}}(\mathbf{\tilde{x}}) = \left( \left(\mathbf{F}_{\mathbf{\theta_a}^{(i)}} * \mathbf{g} \left( \mathbf{\tilde{h}}^{i-1}(\mathbf{\tilde{x}}) \right) \right)_{i = 1}^m ,  \mathbf{\tilde{h}}^m(\mathbf{\tilde{x}})  \right)    
    \end{equation}
    where:
    \begin{itemize}
        \item $\mathbf{F}_{\mathbf{\theta_a}^{(i)}}$ is the filter bank at scale $i$.  $\mathbf{\theta_a}^{(i)}$ are the convolution kernels all of the same square size.
        
        \item $\mathbf{\tilde{h}}: \mathbf{y} \mapsto \mathbf{\bar{u}}( \mathbf{h} * \mathbf{y})$ is a low-pass filtering ($\mathbf{h}$) followed by decimation ($\mathbf{\bar{u}}$). 
        
        \item $\mathbf{g}$ is a high-pass filter given by: $\mathbf{g}(\mathbf{y}) = \mathbf{y} - \mathbf{u}(\mathbf{\tilde{h}}(\mathbf{y}))$, where $\mathbf{u}$ is the upsampling operation performed using a bicubic interpolator.
    \end{itemize}
    Only $\mathbf{F}_{\mathbf{\theta_a}}^{(i)}$ filters are learned, while keeping the high and low pass filters $(\mathbf{g}, \mathbf{h})$ fixed.

    \item $\mathbf{T}_{\mathbf{\theta_t}}$ is the thresholding function. \cite{ramzi:hal-03346892} chose the soft-thresholding operation since it offers more stability \citep{abrial2007}. Owing to the linear nature of the analysis operator, one could straightforwardly apply the thresholding strategy to any other type of noise, including non-stationary Gaussian noise. In case of white Gaussian noise with variance $\sigma^2$, this function is described as:
    \begin{equation}
    \label{thresh}
        \mathbf{T}_{\mathbf{\theta_t}} \left( \left( (\mathbf{d}_i)_{i = 1}^m , \mathbf{c} \right) , \sigma \right) = \left( \left( \left( t_{ij}(d_{ij}, \sigma) \right)_{i = 1}^{J_i}\right)_{i = 1}^m , \mathbf{c} \right)
    \end{equation}
    where $t_{ij}(\mathbf{d}, \sigma) = \hat{\sigma}_{ij}  ST \left(\frac{1}{\hat{\sigma}_{ij}} d_{ij}, \theta_{T}^{(ij)} \sigma \right)$, with:
    \begin{itemize}
        \item $d_{ij} \in \mathbb{R}^{ \frac{n}{2^{i-1}} \times \frac{n}{2^{i-1}}}$ the output of the $j^{th}$ filter of $i^{th}$ scale.
        \item $\hat{\sigma}_{ij}$ the estimated standard deviation of $d_{ij}$ when white Gaussian noise with variance $1$ is input to the transform.
        \item $\theta_{T}^{(ij)}$ the level of thresholding applied to the $j^{th}$ analysis filter at scale $i$.
        \item $ST(\mathbf{d}, \mathbf{s})$ the soft-thresholding function applied point-wise on $\mathbf{d}$ with threshold $\mathbf{s}$: $ST(\mathbf{d}, \mathbf{s}) = \text{sign }(\mathbf{d})\max(|\mathbf{d}| - \mathbf{s}, 0)$.
    \end{itemize}

\item $\mathbf{S}_{\mathbf{\theta_s}}$ is the synthesis function, which is equivalent to the wavelet reconstruction operator with learned filters and is linear. It is described recursively as:
\begin{equation}
   \mathbf{S}_{\mathbf{\theta_s}} \left(  (\mathbf{d}_i)_{i = 1}^m , \mathbf{c} \right) = \mathbf{S}_{\mathbf{\theta_s}}^{(m-1)} \left(  (\mathbf{d}_i)_{i = 1}^{m-1} , \mathbf{u}(\mathbf{c}) + \mathbf{F}_{\mathbf{\theta_s}^{(m)}} * \mathbf{d}_m \right)
\end{equation}
where $\mathbf{S}_{\emptyset}( \emptyset, \mathbf{c}) = \mathbf{c}$ and:
\begin{itemize}
    \item $\mathbf{F}_{\mathbf{\theta_s}^{(i)}}$ is the filter bank at scale $i$. $\mathbf{\theta_s}^{(i)}$ are the convolution kernels all of the same square size.
    \item $\mathbf{u}$ is an upsampling operation performed by a bicubic interpolator. 
\end{itemize}
\end{enumerate}

\begin{figure}[h!]
\begin{center}
\includegraphics[width=\columnwidth]{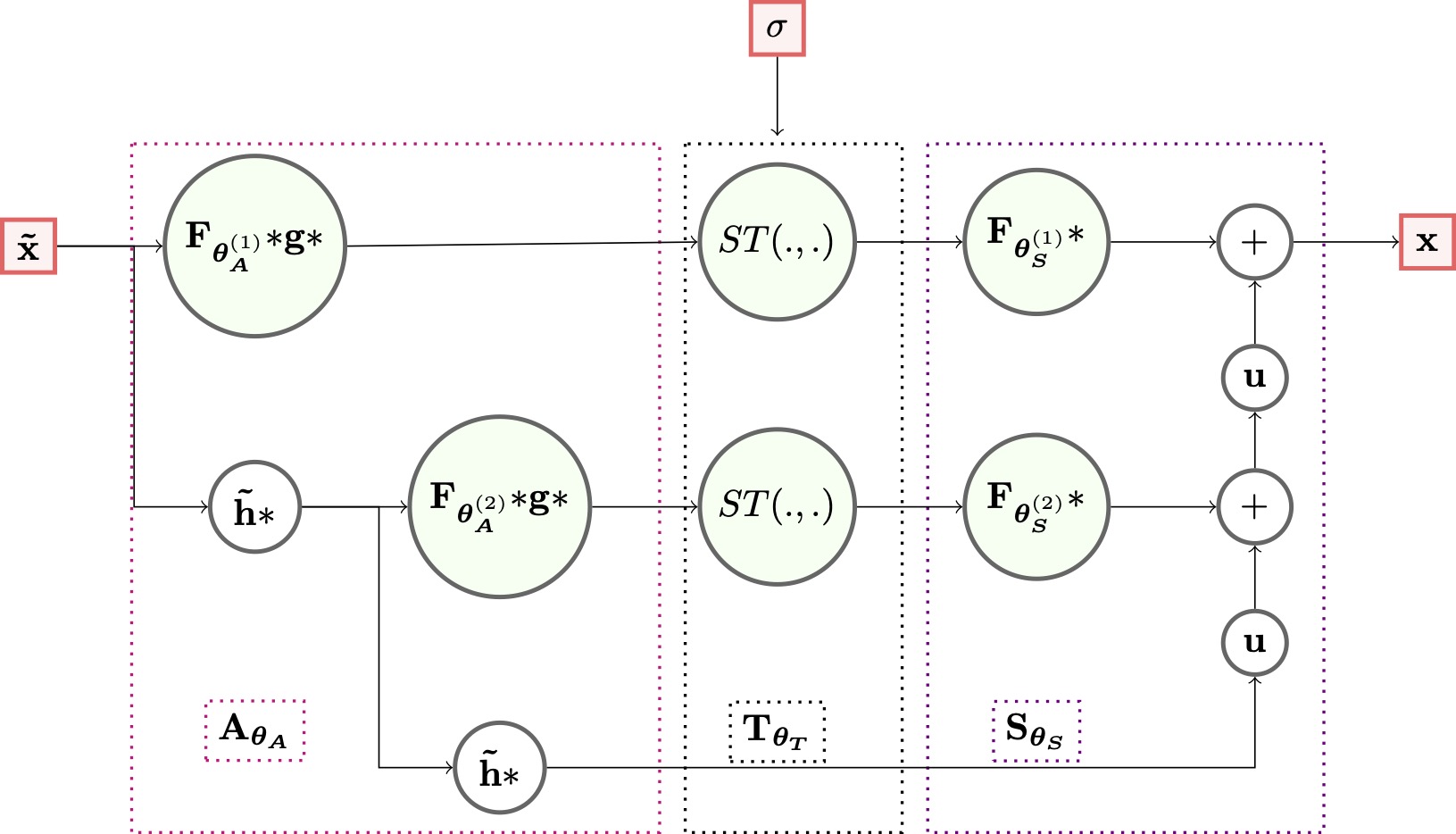}
\end{center}
\caption{\label{fig:learnlets-schema} Schematic representation of the Learnlet model, with $m=2$ scales. Inputs/outputs are represented by the red nodes. The light-green nodes denote functions with learnable parameters. Credit: \cite{ramzi:hal-03346892}.}
\end{figure}

To make Learnlets behave as closely as possible to wavelets and hence more intuitive, \cite{ramzi:hal-03346892} constrain the thresholding levels to lie within $[0, 5]$ and force the analysis filters to have a unit norm. Figure~\ref{fig:learnlets-schema} shows a schematic of the Learnlet architecture . The full Python implementation of the code is publicly available. \footnote{https://github.com/zaccharieramzi/understanding-Unets}.

\subsection{Deconvolution with Learnlets}

Similar to Tikhonet deconvolution, in Learnlet deconvolution, we find the closed-form solution $\hat{\mathbf{x}}$ as given in \ref{eq:tikhonov}. $\hat{\mathbf{x}}$ is then post-processed by a Learnlet network, which is also trained on pairs of Tikhonov outputs $\hat{\mathbf{x}}$ and ground truth images $\mathbf{x_t}$ to minimize the MSE (equation \ref{eq:learning}). The following model parameters were chosen for the Learnlet architecture, which amounted to $372,000$ trainable network parameters \citep{ramzi:hal-03346892}:
\begin{itemize}
    \item $m = 5$ scales.
    \item $256$ learnable analysis filters plus $1$ fixed identity analysis filter $\mathbf{F}_{\mathbf{\theta_a}^{(i)}}$ of dimensions $3 \times 3$.
    \item $257$ learnable synthesis filters $\mathbf{F}_{\mathbf{\theta_s}^{(i)}}$ of dimensions $5 \times 5$.
\end{itemize}

Additionally, Learnlets also require the standard deviation of the noise $\mathbf{\sigma_{noise}}$ as an input to the model. An interesting thing to note here is that after the Tikhonov deconvolution step, the nature of the noise changes from White Gaussian (uncorrelated) to Colored Gaussian (correlated). Even though \cite{ramzi:hal-03346892} developed Learnlets for White Gaussian Noise (WGN) removal, we observe that the algorithm is able to adapt its thresholding step (equation \ref{thresh}) and efficiently recover the solution. This is supported by the results shown in section \ref{results}. Note that there also exist methods to estimate the noise standard deviation in an image with an accuracy of around $1\%$ \citep{starck2015sparse}.

In all of the deconvolution methods -
\begin{itemize}
    \item The Tikhonov deconvolution step remains the same, while the network that post-processes the Tikhonov output $\mathbf{\hat{x}}$ differs.
    \item The success lies on the training accuracy between the Tikhonov deconvolution $\mathbf{\hat{x}}$ and the ground truth image $\mathbf{x_t}$.
    \item The network used for post-processing acts as a denoiser. The efficiency and accuracy of the denoising step governs the reconstructed output, which should ideally be as close to the target image $\mathbf{x_t}$ as possible.
\end{itemize}

\section{Dataset \& Experiments}
\label{exp}

The code was implemented in Python 3.6, using TensorFlow 2.2 \citep{tensorflow2015-whitepaper} for model design. All simulations and trainings were performed on the Yggdrasil supercomputer based at the University of Geneva, using a single Titan RTX Turing GPU with 24GB RAM for each job. 

\subsection{Dataset Generation}
\label{datagen}

\subsubsection{Ground Truths - CANDELS}
The Cosmic Assembly Near-IR Deep Extragalactic Legacy Survey, CANDELS \citep{CANDELS, CANDELS_HST}, was conceived to record the first third of galactic evolution from $z = 8$ to $1.5$ via deep imaging of more than 250,000 galaxies. The entire survey consists of five different image mosaics (GOODS-N, GOODS-S, EGS, UDS, COSMOS), each covering different regions of the sky. To generate our training dataset, using Python, we extracted cutout windows of dimensions $128 \times 128$ pixels from the CANDELS FITS image mosaics in the $F606W$ filter (V-band) by centering them at the object centroid, without any dynamic range-scaling. For larger objects, we adaptively increased the cutout window dimensions to completely enclose them. In that case, if the window dimensions exceeded $128 \times 128$ pixels, we downsampled the images to $128 \times 128$. To select good galaxy candidates and exclude point-sized objects, we use the following filtering criteria:
\begin{itemize}
    \item MAG\_AUTO $< 26$ (AB magnitude in SExtractor “AUTO” aperture) 
    \item $\text{Flux\_Radius}_{80} > 10$ ($80\%$ enclosed flux radius in pixels) 
    \item FWHM $> 10$ (full width at half maximum in pixels)
\end{itemize}
In all, we end up with $22,317$ ground-truth images. 

\subsubsection{Simulations}
\label{datasim}
All the extracted ground truth images are convolved with a Gaussian PSF with an FWHM of 15 pixels, which visually blurs the images such that individual small-scale structures are lost. As seen from equation \ref{eq:tikhonov}, since the algorithm takes the PSF as an input parameter and performs Tikhonov deconvolution in the first step, the method would work with any other kind of blurring kernel. The actual performance of the method is mainly dependent on the neural network training, which corresponds to the denoising step. After convolution, we add white Gaussian noise with a standard deviation $\mathbf{\sigma_{noise}}$ having a value such that the faintest object in our dataset has a peak SNR close to 1 and is hence barely visible. For this value of $\mathbf{\sigma_{noise}}$, we observe a range of SNR values depending on the magnitude of the galaxy. Eventually, the $\mathbf{\sigma_{noise}}$ values are also required as inputs to the Learnlet architecture. Finally, the batch of noisy simulations and their corresponding ground truth images is randomly split into Train-Test subsets in the ratio $0.9:0.1$

\subsection{Pre-processing}
\label{preprocessing}
Each image $\mathbf{x}_{(i)}$ was normalised by subtracting its own mean $\mu_{(i)}$ and scaling within the $[-1,1]$ range as follows: 
\begin{equation}
   \frac{\mathbf{x}_{(i)} - \mu_{(i)}}{\text{max }[\mathbf{x}^t_{(i)} - \mu^t_{(i)}]}    
\end{equation}

where $\mathbf{x}^t_{(i)}$ and $\mu^t_{(i)}$ denote the $i^{th}$ target image and its corresponding mean. The denominator ensures that all target images have a maximum pixel intensity of 1, and that all noisy images are scaled with respect to their corresponding targets such that the ratios of the peak intensities between them remain the same before and after normalization. Since the flux of the target image is spread out by the PSF as a result of convolution, the noisy images have a lower peak intensity than their corresponding target images. One of the main goals of astrophysical deconvolution is to recover this drop in peak intensity. During training, we augment our dataset by performing random rotations in multiples of 90°, translations and flips along horizontal \& vertical axes. Data augmentation is also beneficial for making neural networks invariant to rotations and translations.

\subsection{Training}
\label{training}
After performing a Tikhonov deconvolution as described by equation \ref{eq:tikhonov}, the networks were trained to learn the mapping from the Tikhonov outputs $\hat{\mathbf{x}}$ to the ground truth images $\mathbf{x_t}$ while minimizing the MSE (equation \ref{eq:learning}). The images were fed to the networks in mini-batches of size $32$. All networks were trained until convergence was achieved. We trained the networks with an Adam optimizer \citep{kingma2014adam} and and an initial learning rate of $10^{-3}$, which was then decreased by half every 25 epochs until it reached a minimum of $10^{-5}$. The way we generate our dataset (as described in section \ref{datagen}) results in images with varying noise levels, which ensures that the training is not biased by a certain noise level. Table \ref{tab:perf} shows a performance comparison of the three methods.

\begin{table}[htbp]
\caption{Performance comparison of the three deconvolution methods. The runtime per image was calculated on the same GPU on which the networks were trained.}
\begin{center}
\begin{tabular}{|c|c|c|c|c|}
\hline
\textbf{\textbf{Method}} & \textbf{\textbf{No. of Parameters}} & \textbf{\textbf{Epochs}} & \textbf{\textbf{Training Time (hrs.)}} & \textbf{\textbf{Runtime per image (ms)}} \\
\hline
Tikhonet & $184,301$ & $40$ & $3.65$ & $4.02$\\
Learnlet & $372,000$ & $150$ & $9.81$ & $30.8$\\
Unet-64 & $31,023,940$ & $500$ & $16.9$ & $26.3$\\
\hline
\end{tabular}
\label{tab3}
\end{center}
\label{tab:perf}
\end{table}

\section{Results}
\label{results}

\begin{figure}[h!]
\begin{center}
\includegraphics[width=\columnwidth]{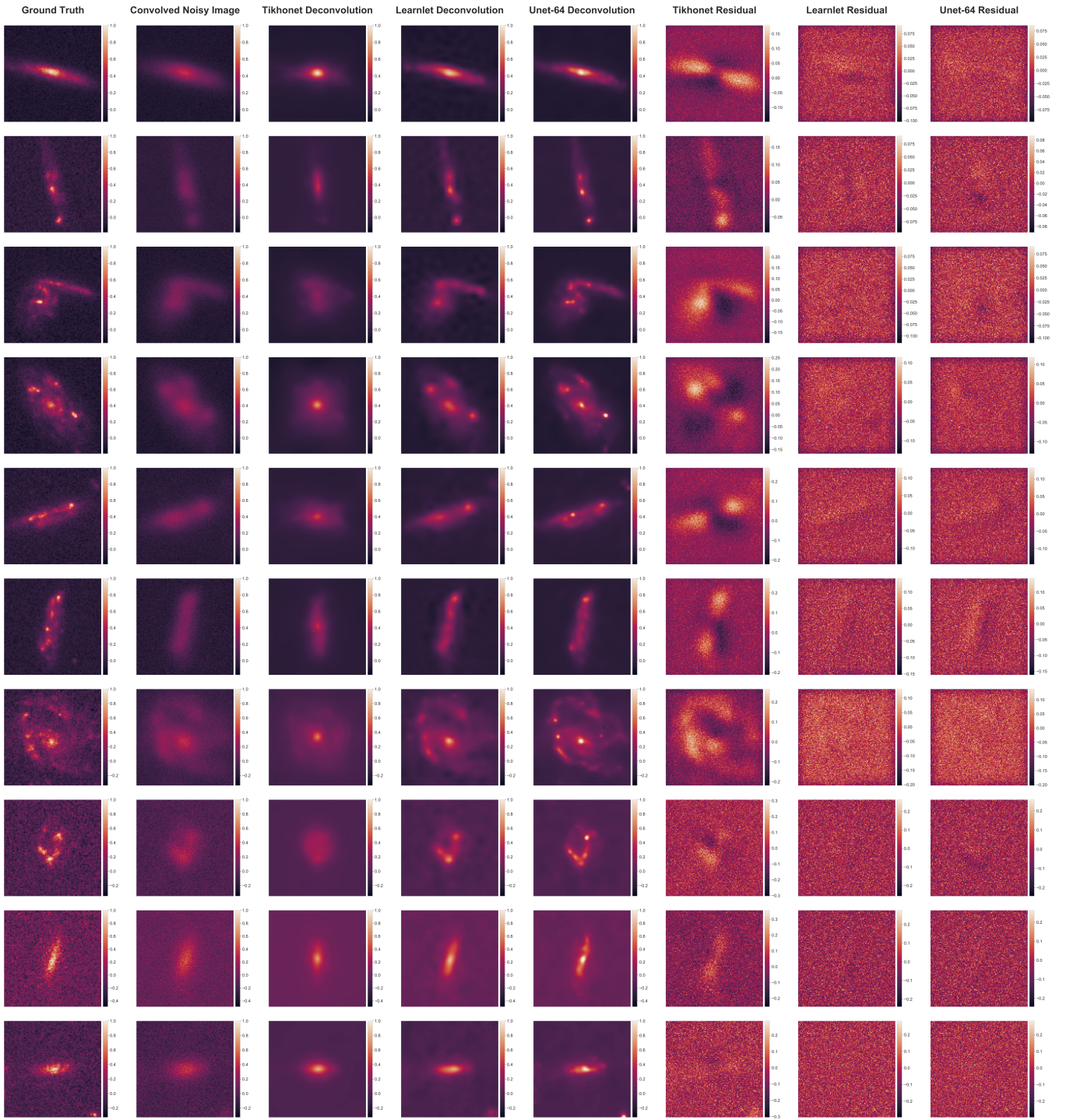}
\end{center}
\caption{Ten examples of deconvolved images in decreasing order of SNR along with the corresponding residuals. Unet-64 results in the best reconstructions with the cleanest residuals.}\label{fig:deconv_outputs} 
\end{figure}

For evaluation, our test dataset contains images with varying noise levels (\ref{datasim}). To perform a quantitative comparison, we use the Normalised Mean Squared Error (NMSE) and the Structural Similarity Index Measure (SSIM), two commonly used metrics to quantify image-reconstruction quality. To measure the NMSE and SSIM, we weighted all images by a Gaussian window centered around the galaxy. The FWHM for each object was obtained from its catalog and used for the Gaussian weighting, thus ensuring that the window would only encircle the object and discard the noise present in the background of the target images. An illustration of the same along with the noisy Tikhonov output $\hat{\mathbf{x}}$ is shown in figure \ref{fig:tikho_win}. To visually analyze the final deconvolution results, in figure \ref{fig:deconv_outputs}, we show ten examples of reconstructed galaxies with different morphologies in decreasing order of SNR. We observe that Tikhonet is mostly able to reconstruct pixels around the central region. Consequently, it is able to deconvolve small objects but is unable generalize well to extended objects. Learnlet and Unet-64, on the other hand, are also able capture small-scale structures and non-central peaks present in the images, while well-preserving the global shape simultaneously. Moreover, visually, their performace is consistent on both high and low SNR images, with an even further improvement for Unet-64. We also compute and show the residual maps for these three deconvolution methods. The residuals are defined as:

\begin{figure}[h!]
\begin{center}
\includegraphics[width=\columnwidth]{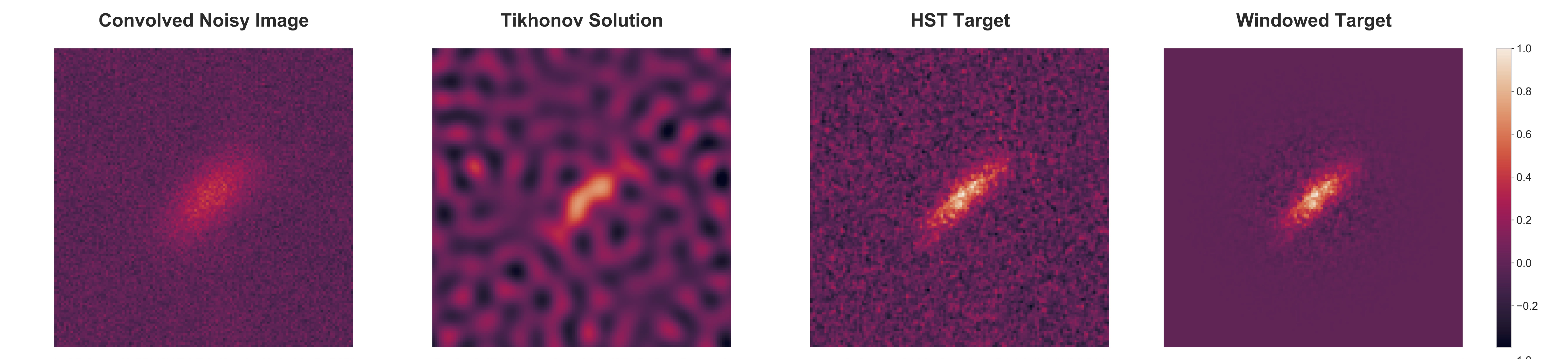}
\end{center}
\caption{An example of Tikhonov output (second image) along with the weighted Target image using a Gaussian window (fourth image).}\label{fig:tikho_win}
\end{figure}

\begin{equation}
    \label{eq:residual}
    \text{Residual} = \mathbf{y - h \ast N_{\theta}(\hat{x})}
\end{equation}

where $\mathbf{y}$ corresponds to the noisy image, $\mathbf{h}$ is the PSF, $\hat{\mathbf{x}}$ is the noisy Tikhonov input, and $\mathbf{N_{\theta}}$ is the network model. 

Ideally, one should see pure noise in the residuals. We can see that there is still some structure present in the Tikhonet residuals, which comes as a result of incomplete reconstruction. The residuals are more structured in case of extended objects as compared to smaller ones, which implies that Tikhonet struggles to  recover extended objects. Contrastingly, the Learnlet and Unet-64 residuals contain minimal structure with no orientational preference, which indicates that the reconstruction is better regardless of the size of the galaxy.

\begin{figure}[h!]
\begin{center}
\includegraphics[width=\columnwidth]{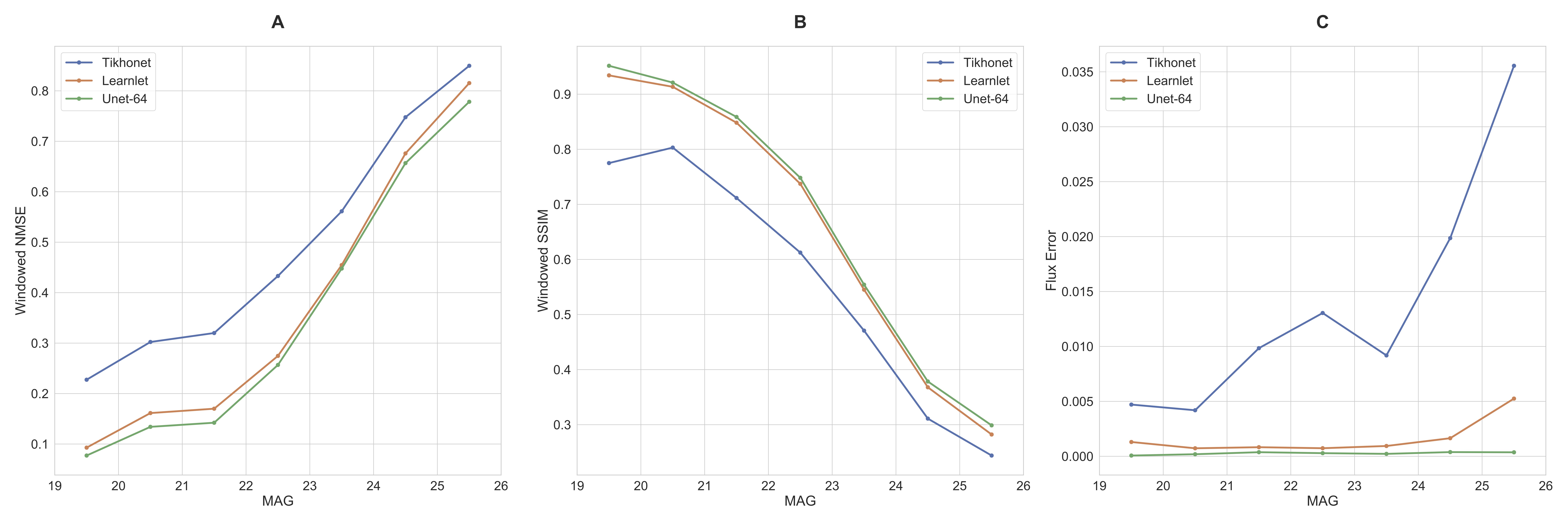}
\end{center}
\caption{\textbf{(A)} Windowed NMSE as a function of the Magnitude (MAG), \textbf{(B)} Windowed SSIM as a function of MAG. An SSIM of 1 implies that the images are identical. \textbf{(C)} The Flux Error as a function of MAG. Unet-64 generalizes very well to the entire range of magnitudes.}\label{fig:NMSE_SSIM} 
\end{figure}

In figure \ref{fig:NMSE_SSIM}, we show the trends obtained for windowed NMSE and SSIM on our test dataset images binned according to their magnitudes. All networks perform better on low magnitude (or high SNR) images, and there is a decrease in performance with an increase in the magnitude (or decrease in SNR). Furthermore, for the range of magnitudes and metrics considered, Unet-64 outperforms all other methods, followed by Learnlet. The mean NMSE improvement is $11.4\%$ when going from Tikhonet to Learnlet, and $2.2\%$ when going from Learnlet to Unet-64. Similarly, the mean SSIM improvements are $10\%$ and $1.2\%$ respectively. We also compare the three networks based on their ability to preserve the image flux, which is essentially the sum of all pixel intensities. As seen in figure \ref{fig:NMSE_SSIM}, although the errors are very low for all the networks, Unet-64 performs the best in terms of flux preservation, followed by Learnlet and Tikhonet. More specifically for Tikhonet, the error steeply increases with an increase in magnitude. On the other hand, Unet-64 and Learnlet have flatter curves and generalize well to all magnitudes, with Unet-64 generalizing the best. Based on the metrics considered, we finally conclude that Unet-64 deconvolution surpasses Learnlet and Tikhonet deconvolution.

\section{Conclusion}
\label{conclusion}
We proposed a new deconvolution method based on the Learnlet transform. Consequently, we compared the performance of Tikhonet, Learnlet, and Unet-64 for image deconvolution in the astrophysical domain by adapting a two-step approach involving a Tikhonov deconvolution followed by post-processing with a denoiser. Visually, we observed that Unet-64 and Learnlet are able to capture the small-scale structures in the images in addition to the global shape, while simultaneously well-preserving the flux. Since the networks were evaluated on a range of noise levels, we could conclude that Unet-64 and Learnlet generalize well unlike Tikhonet, with Unet-64 having the best performance. These observations are further supported by the quantitative metrics used for comparison, where Unet-64 outperforms Learnlet and Tikhonet. As seen in Table \ref{tab:perf}, although Tikhonet has the smallest runtime, it does not perform well on extended objects. Furthermore, Unet-64 has a smaller runtime per image compared to Learnlet, making it more efficient. This makes it an ideal candidate to be used for astrophysical image deconvolution. The strategy we proposed in this paper for building the training dataset seems very efficient and could easily be applicable to Euclid images by using the deep field survey to build a training dataset that could eventually be used for the wide field survey. More experiments will however be required to check more in details the generalization properties.

It would also be interesting to compare all the networks with the shape constraint investigated in \cite{shapenet}. For Learnlets, as in the case of wavelets, various different kinds of noise could be considered on a single model with an undecimated implementation by adjusting the thresholding function according to the noise. One could enhance the Unet-based methods by substituting the Unet with a sophisticated denoiser like the Deep Iterative Down-Up CNN (DIDN) \citep{yu2019deep}, and adopt a similar strategy for Learnlets by integrating them as elementary units of the DIDN model. On a slightly different note, it would also be interesting to test the concept of unrolled networks \citep{monga_unroll} and neural augmentation \citep{behrens2021neurally} in order to mimic iterative deconvolution algorithms with theoretical guarantees. A powerful and accessible deconvolution method would allow for the reconstruction of a cleaner estimation of the sky with less data than classical methods, which is essential for optimising the observing time and the amount of data required to attain a given image reconstruction fidelity \citep{shapenet}. On a broader scope, these deconvolution methods could also be applied to other fields.

\section*{Conflict of Interest Statement}
The authors declare that the research was conducted in the absence of any commercial or financial relationships that could be construed as a potential conflict of interest.




\section*{Acknowledgments}
This research was funded by the Swiss National Science Foundation grant number CRSII5\_198674. This work is based on observations taken by the CANDELS Multi-Cycle Treasury Program with the NASA/ESA HST, which is operated by the Association of Universities for Research in Astronomy, Inc., under NASA contract NAS5-26555.

\bibliographystyle{Frontiers-Harvard} 
\bibliography{bibliography.bib}

\end{document}